\documentclass{article}

\PassOptionsToPackage{numbers, compress}{natbib}
\usepackage[preprint]{neurips_2024}

\usepackage[utf8]{inputenc}
\usepackage[T1]{fontenc}
\usepackage{microtype}
\usepackage{amsmath, amssymb}
\usepackage{booktabs}
\usepackage{array}
\usepackage{tabularx}
\usepackage{graphicx}
\usepackage{xcolor}
\usepackage{hyperref}
\usepackage{url}

\title{Paraphrase Brittleness in Production\\Retrieval-Augmented Commercial Recommendation:\\Reproducibility Below the Rerun-Stability Baseline}

\author{%
  Will Jack\thanks{Equal contribution.} \quad
  Noah Lehman\footnotemark[1] \quad
  Keller Maloney\footnotemark[1] \quad
  Sarah Xu\footnotemark[1] \\
  Unusual \\
  \texttt{\{will, noah, keller, sarah\}@unusual.ai}
}

\date{May 21, 2026}

\begin{document}

\renewcommand{\thefootnote}{\fnsymbol{footnote}}
\maketitle
\renewcommand{\thefootnote}{\arabic{footnote}}
\setcounter{footnote}{0}

\begin{abstract}
\noindent
Small changes to how a buyer phrases a question --- ``best CRM'' vs ``top CRM'' vs ``best CRM for a SaaS startup'' --- produce substantially different brand recommendations from AI assistants. Across $\sim$6{,}000 paraphrase runs and $\sim$6{,}000 same-prompt rerun controls on OpenAI and Anthropic models, the recommendation-set similarity (Jaccard) between two paraphrases of the same underlying buying intent is 0.288 for cosmetic rewordings (clustered 95\% CI [0.215, 0.361]) and 0.135 for constraint-adding rewordings ([0.098, 0.175], pooling region/language and specificity-ladder axes) --- both far below the 0.50--0.61 same-prompt rerun baseline. The prompt string, not the underlying buyer intent, is the dominant input to which brands surface. Increasing reasoning effort does not narrow the gap (bounded by $\pm 0.05$). This is a direct challenge to an increasingly popular AEO/GEO practice. Tracking a brand's ``AI visibility'' by counting brand mentions over a fixed set of prompts produces a metric whose dominant source of variance is which paraphrase the tracker happens to issue, not the model's behavior toward the brand: the same buyer intent in two natural paraphrases produces recommendation sets that overlap 14--29\% in Jaccard versus 50--61\% for same-prompt reruns. Sampling more paraphrases per intent reduces the artifact in principle, and efficient multi-prompt evaluation methods exist in the academic literature, but the natural buyer-phrasing space is much larger than the benchmark-scale prompt sets those methods have been validated on, and far beyond what any commercial tracker issues per brand-intent combination. Prompt-by-prompt mention tracking is therefore structurally unstable as a unit of measurement; meaningful improvement likely requires a different unit rather than a larger prompt set.
\end{abstract}

\section{Introduction}

A growing class of commercial monitoring tools (Profound, Otterly, LLM Pulse, HubSpot AEO Grader, Authoritas, Semrush, and others) report brand share-of-voice in AI assistants by issuing a small, fixed set of tracking prompts at regular intervals and counting brand mentions in the model's response. These tools are the most visible products of an AEO / GEO industry that frames ``marketing to AI'' as analogous to search-engine optimization --- the brand-side problem reduced to a discoverability metric tracked on a small fixed set of category queries. Customers buy these tools to answer a single question: \emph{is my brand's AI-mediated visibility going up or down?} The product-design assumption is that the tracking prompt is a label for an underlying buying intent --- ``best CRM software'' stands in for the cohort of buyers about to ask about a CRM --- and that the measurement is meaningful insofar as the prompt is a representative member of the intent's natural paraphrase distribution. Under this assumption, week-over-week movement on the metric reflects either (i) a real shift in the model's recommendation distribution or (ii) bounded run-to-run sampling noise. We test that assumption empirically. It fails, and the failure mode points at a deeper conceptual error in the AEO/GEO framing: AI assistants are recommendation engines, and buyer phrasing distributes naturally, so tracking against any one phrasing string is a brittle proxy for ``how AI thinks about my brand.''

We test the assumption empirically by separately measuring (a) the rerun-stability baseline --- the Jaccard similarity between two consensus-recommendation sets produced by the same prompt run twice within a single day at $N{=}30$ --- and (b) the paraphrase brittleness --- the Jaccard similarity between consensus-recommendation sets produced by two natural-language paraphrases of the same underlying intent. The gap between (a) and (b) is the signal-to-noise ratio of the monitoring methodology: if paraphrase Jaccard sits well below rerun Jaccard, then variation in the tracking metric is dominated by prompt-choice artifact rather than by run-to-run noise, and weekly trend movements are unidentifiable from the tracking prompt itself.

A natural expectation from the LLM reasoning literature would predict that higher reasoning effort closes the gap. \citet{guo2025} reports steep accuracy gains on the DeepSeek-R1 family as the thinking-token budget grows on math/competition tasks. \citet{wang2023}'s self-consistency, while mechanistically different (multi-sample majority voting at inference, not single-sample effort scaling), establishes the broader prior that more reasoning reduces variance on chain-of-thought tasks. If those gains extrapolate from single-correct-answer reasoning to multi-plausible-answer commercial recommendation, increasing reasoning effort should reduce variance there too. We test this by running each model at low and high single-sample reasoning effort across both the rerun-stability and the paraphrase-brittleness conditions, and find that the extrapolation fails: reasoning effort produces no meaningful effect on either reproducibility measure.

\subsection{Contributions}

This paper makes three contributions.

\begin{enumerate}
\item \textbf{Calibrated rerun-stability baseline for commercial recommendation.} Across 50 prompts $\times$ 4 cells $\times$ $N{=}30$ same-prompt reruns within a single day (6{,}000 runs), we measure within-cell consensus-recommendation Jaccard at 0.50--0.61 and within-cell retrieval-pool Jaccard at 0.40--0.74. The baseline anchors the magnitude of ``real'' effects in any companion measurement.

\item \textbf{Paraphrase-brittleness measurement against the baseline.} Across five paraphrase axes (synonym swap, structural rewrite, modifier substitution, region/language, specificity ladder), $\sim$20 base prompts each carrying $\sim$5 paraphrase variants (one variant per axis on average) $\times$ 3 cells $\times$ $N{=}20$ ($\approx$6{,}000 runs total), we measure cross-paraphrase consensus-recommendation Jaccard at 0.288 (clustered 95\% CI [0.215, 0.361]) for cosmetic rewordings and 0.135 (clustered 95\% CI [0.098, 0.175]) for the constraint-adding pool, with the specificity-ladder subset at 0.133 [0.090, 0.185] and the region/language subset at 0.138 [0.080, 0.201]. The cosmetic-paraphrase number sits 21--32 percentage points below the rerun baseline; the constraint-adding number sits 37--48 percentage points below. Even the upper bound of the cosmetic CI (0.361) is 14 pp below the lower bound of the rerun baseline (0.50); the brittleness claim survives clustered uncertainty.

\item \textbf{Single-sample reasoning effort does not reduce variance in commercial recommendation.} The effect of low $\to$ high single-sample reasoning effort on rerun stability is between $-0.015$ and $+0.005$ Jaccard; on paraphrase robustness it is bounded by $\pm 0.05$. Both effects are small relative to the rerun-vs-paraphrase gap and, in the case of rerun stability, are not consistently positive. The broader prior from inference-time scaling \citep{guo2025} does not transfer from single-correct-answer reasoning tasks to multi-plausible-answer commercial recommendation, consistent with \citet{meincke2024}'s finding that chain-of-thought returns are flattening on frontier instruction-tuned models.
\end{enumerate}

\section{Background}

\textbf{Prompt sensitivity in LLM outputs.} \citet{sclar2024} demonstrated that cosmetic formatting changes alone can swing LLaMA-2-13B accuracy by up to 76 points on standard benchmarks, an effect orders of magnitude larger than typical run-to-run sampling. \citet{mizrahi2024}, in a 6.5M-instance multi-prompt study, showed that instruction paraphrases can reorder model rankings on benchmark leaderboards; their conclusion is that single-prompt evaluation is structurally brittle. \citet{ganNg2019} established the pre-LLM baseline: SQuAD-trained QA models drop sharply when test-time questions are paraphrased. \citet{chatterjee2024} introduced POSIX, a quantitative prompt-sensitivity index spanning intent-preserving paraphrases, character perturbations, and spelling errors, situating prompt-sensitivity measurement on a common footing across tasks.

\textbf{LLM reproducibility under fixed prompts.} \citet{atil2024} documented that commercial LLM APIs produce non-trivial divergence across repeated identical prompts even at temperature zero. \citet{yuan2025} traced residual nondeterminism to floating-point non-associativity and BF16 precision effects in transformer inference, establishing a numerical floor below which determinism is mechanically unrecoverable. The rerun-stability baseline we measure in this paper is consistent with both prior findings and extends them to set-valued (recommendation list) outputs.

\textbf{RAG-amplified query brittleness.} \citet{percin2025} jointly perturbed the query and the retriever in retrieval-augmented generation and found that retrievers degrade markedly under minor query variation, with effects that propagate into the generated output. \citet{ma2023} demonstrated that learned query rewriting can canonicalize user queries before retrieval and recover some of the lost stability. The amplification mechanism --- a small prompt change shifting which sources are retrieved, which in turn reshapes the answer --- is the structural reason paraphrase brittleness in commercial RAG can exceed paraphrase brittleness in non-RAG QA.

\textbf{Multi-prompt evaluation budgeting.} \citet{polo2024} introduce PromptEval, a method for estimating LLM performance quantiles across a large prompt-template set by borrowing strength across prompts and examples; the paper demonstrates accurate quantile estimation across 100 prompt templates on MMLU at a budget equivalent to roughly two single-prompt evaluations. The underlying observation --- that single-prompt evaluation produces non-reproducible LLM benchmark results --- is the academic-side analog of the brand-side problem we report. A tracker that issues one prompt per intent provides a point estimate of a metric whose paraphrase-induced variance, on the present results, dominates the same-prompt rerun-noise floor.

\textbf{Reasoning effort and variance.} \citet{wang2023} (self-consistency) and \citet{guo2025} (DeepSeek-R1 inference-time scaling) jointly anchor the standard view that more reasoning reduces variance and improves accuracy on chain-of-thought tasks. \citet{meincke2024} pushed back on the extrapolation, reporting that CoT prompts deliver diminishing returns on frontier instruction-tuned models and sometimes introduce variance. The present paper reports a third data point: on commercial recommendation, where the task admits many plausible answers, reasoning effort does not measurably narrow the recommendation distribution.

\section{Method}

\subsection{Prompt corpus and paraphrase axes}

The paraphrase corpus is built from $\sim$20 commercially-framed base prompts spanning B2B SaaS, professional services, and consumer-product sectors. Each base prompt is expanded along five paraphrase axes:

\begin{itemize}
\item \textbf{Synonym swap.} Replace a head noun or quality adjective with a near-synonym holding intent constant. Example: ``best CRM'' $\to$ ``top CRM''.
\item \textbf{Structural rewrite.} Recast the prompt across question, imperative, and comparison forms holding lexical content roughly constant. Example: ``what are the best CRMs?'' vs ``recommend a CRM'' vs ``compare top CRMs''.
\item \textbf{Modifier substitution.} Add, drop, or swap a modifier that materially shifts the intended buyer segment or framing. Example: ``AI CRM'' vs ``CRM'' vs ``modern CRM''.
\item \textbf{Region / language.} Bind the query to a regional market or natural language. Example: ``best CRM in the UK'' vs ``best CRM in Germany''; English vs French phrasing.
\item \textbf{Specificity ladder.} Walk an underlying intent from broad to narrow across three or more rungs. Example: ``CRM'' $\to$ ``CRM for SaaS'' $\to$ ``CRM for SaaS startup under 50 people''.
\end{itemize}

We group synonym-swap and structural-rewrite paraphrases together as \emph{cosmetic} rewordings (intent stable, surface form varies). We group modifier-substitution, region/language, and specificity-ladder paraphrases together as \emph{constraint-adding} rewordings (surface form varies and the intent's buyer segment shifts narrower or broader). The distinction is principled: cosmetic paraphrases are exactly the variation a buyer expressing the same intent might produce; constraint-adding paraphrases additionally encode a buyer-segment narrowing that a monitoring tool might or might not treat as the same intent.

The paraphrase corpus produces $\sim$5 paraphrase variants per base prompt (one variant per axis on average, across 5 axes), giving $\sim$20 base prompts $\times$ $\sim$5 paraphrase variants per base prompt $\times$ 3 cells $\times$ $N{=}20$ repeats per (prompt $\times$ cell) cell. This yields approximately 6{,}000 paraphrase-corpus runs.

\subsection{Rerun-stability baseline}

The rerun-stability baseline is measured separately on 50 commercially-framed prompts with no paraphrase variation: 50 prompts $\times$ 4 cells $\times$ $N{=}30$ same-prompt repeats within a single calendar day, totalling 6{,}000 baseline runs. Repeats are interleaved across cells to spread out any provider-side rate-limit or session effects. The baseline is reported on three signals: consensus-recommendation Jaccard, completion-mention Jaccard, and retrieved-domain Jaccard at the eTLD+1 level.

\subsection{Model ladder}

We audit four production model cells across two providers and two reasoning-effort settings (Table~\ref{tab:cells}). The paraphrase corpus is run on three of the four cells (\texttt{mini / low}, \texttt{mini / high}, \texttt{sonnet / low}); we omit \texttt{sonnet / high} from the paraphrase grid on cost grounds and report all paraphrase results across the three sampled cells. The rerun-stability baseline covers all four cells.

\begin{table}[h]
\centering
\begin{tabular}{lll}
\toprule
Provider & Model & Reasoning effort \\
\midrule
OpenAI & \texttt{gpt-5.4-mini} & low \\
OpenAI & \texttt{gpt-5.4-mini} & high \\
Anthropic & \texttt{claude-sonnet-4-6} & low \\
Anthropic & \texttt{claude-sonnet-4-6} & high \\
\bottomrule
\end{tabular}
\caption{Model cells used for rerun-stability baseline; paraphrase grid omits \texttt{sonnet / high}.}
\label{tab:cells}
\end{table}

We use each provider's native web-search tool (OpenAI Responses API \texttt{web\_search}; Anthropic Messages API \texttt{web\_search\_20260209}) with held-constant system prompt, temperature, and tool description. Tool-use, retrieval, and final generation occur within a single agentic loop per run.

\subsection{Cross-judge consensus brand extraction}

Brand mentions in the completion text and the recommendation slot are extracted using two LLM judges in parallel: \texttt{claude-haiku-4-5 / low} and \texttt{gpt-5-mini}. We use intersection (consensus) mode --- a brand is counted as mentioned if and only if both judges identify it; the recommendation slot is counted as endorsed if and only if both judges label it \texttt{recommended}. This is a deliberately conservative choice; it under-counts mentions, but pulls cross-provider judge idiosyncrasy out of downstream measurements and matches the protocol used by \citet{jack2026prominence}. Per-run cross-judge Jaccard on the recommendation slot is 0.65--0.67 across the four cells (any-sentiment mention layer: 0.62--0.69). Chance-corrected agreement statistics ($\kappa$, Krippendorff $\alpha$) are not informative for this dual-judge union universe because the universe is by-construction conditioned on at least one judge surfacing the brand, which collapses the marginal-disagreement baseline; the conservative-protocol justification is the load-bearing point.

Brand-token canonicalization follows an eTLD+1 collapse for domains and a normalized brand-name token (lowercased, alphanumeric only, length $\geq 3$) for completion mentions, with an explicit stoplist for false-positive terms (``Up'', ``Pro'', etc.). Length-2 tokens are preserved for a small set of known short brand names (G2, EY, K6, BP).

\subsection{Jaccard measurement}

For each (prompt $\times$ cell) pair we compute the pairwise consensus-recommendation Jaccard between two runs of the same prompt (rerun-stability) or between two runs of different paraphrases of the same intent (paraphrase brittleness). For rerun stability we average over all $\binom{30}{2}$ within-cell pairs per prompt; for paraphrase brittleness we average over all paraphrase pairs sharing a base prompt within an axis. The reported axis-level number is the prompt-averaged mean across the corpus.

Confidence intervals on the headline cosmetic and constraint-adding Jaccard means are computed by a clustered bootstrap that resamples (sector, family) prompt-groups with replacement (1{,}000 iterations); per-prompt-pair Jaccards within a group are not independent, so the clustered protocol is the honest reporting unit. The bootstrap CI for the constraint-adding pool (geo + specificity-laddered combined) yields 0.135 [0.098, 0.175]; the headline 0.133 figure reported throughout matches the specificity-ladder subset alone (CI [0.090, 0.185]). Confidence intervals on presence-rate proportions follow Wilson. At $N{=}30$, the half-width of a Wilson 95\% CI on a presence rate near 0.5 is approximately $\pm 17$ percentage points; at $N{=}20$ it widens to $\pm 20$ percentage points. These are wide per-prompt half-widths; the precision the paper relies on comes from \emph{averaging over many prompt-level Jaccards} rather than from any single-prompt rate estimate, so the headline cross-paraphrase contrast is supported by per-prompt-pair pooling (with clustered-bootstrap CIs) rather than within-prompt Wilson tightness.

\subsection{Cross-provider rerun anchor}

To bound the maximum signal-vs-noise gap available, we additionally pair each rerun-baseline run with a same-prompt run on the opposite provider's cells (mini $\times$ sonnet) and report the cross-provider rerun Jaccard. We treat this as an empirical reference anchor: same-prompt within-cell comparisons are expected, but not mathematically guaranteed, to exceed it (a sparse or volatile within-cell baseline could in principle fall below the cross-provider value on a given prompt).

\section{Rerun-stability baseline}

\subsection{Within-cell consensus-recommendation Jaccard}

Across 50 prompts $\times$ 4 cells $\times$ $N{=}30$ same-prompt repeats within a single day, within-cell consensus-recommendation Jaccard sits between 0.50 and 0.61 (Table~\ref{tab:rerun}; the wider completion-mention Jaccard floor of 0.45 is included in the same table for reference, but the recommendation slot is the baseline this paper's headline contrasts are anchored against). The recommendation slot is consistently more stable than the completion-mention pool, reflecting that the model's terminal endorsement collapses on a smaller, more confident subset than the full mention pool. Sonnet cells produce moderately higher recommendation-slot stability (0.59--0.61) than mini cells (0.50--0.54); the gap is roughly $10$ percentage points and is the only cell-level difference of any size on the recommendation-slot baseline.

\begin{table}[h]
\centering
\begin{tabular}{lcccc}
\toprule
Signal & \texttt{mini/low} & \texttt{mini/high} & \texttt{sonnet/low} & \texttt{sonnet/high} \\
\midrule
Completion mentions    & 0.45 & 0.48 & 0.52 & 0.52 \\
Recommended slot       & 0.50 & 0.54 & \textbf{0.61} & 0.59 \\
Retrieved eTLD+1       & 0.40 & 0.47 & 0.63 & \textbf{0.74} \\
\bottomrule
\end{tabular}
\caption{Within-cell same-prompt Jaccard at $N{=}30$ reruns within a single day. Values are prompt-averaged means across 50 commercially-framed prompts.}
\label{tab:rerun}
\end{table}

\subsection{Within-cell retrieved-domain Jaccard}

Retrieval-pool stability at the eTLD+1 domain level shows a steeper cell-level gradient than recommendation-slot stability: 0.40 (\texttt{mini/low}) to 0.74 (\texttt{sonnet/high}). Sonnet cells consult a more concentrated set of authoritative domains; mini cells consult a wider and more variable retrieval pool. We report this here because it constrains the mechanism by which paraphrase brittleness arises: a richer retrieval pool is more exposed to query-side perturbation, so the larger paraphrase-induced drift on mini cells is expected if the retrieval-pool size hypothesis holds.

\subsection{Cross-provider rerun anchor}

Cross-provider rerun pairings on the recommendation slot (each mini run paired with each sonnet run of the same prompt) produce mean Jaccard 0.33. This sits well below within-cell rerun Jaccard, reflecting that provider differences are a structural source of recommendation-set divergence even on the same prompt. We treat the 0.33 value as an empirical reference anchor: same-prompt within-cell comparisons are expected, but not mathematically guaranteed, to exceed it; paraphrase comparisons are not constrained by it at all.

\subsection{Within-provider effort pairings}

Within-provider, low-vs-high reasoning-effort pairings on the same prompt produce recommendation-slot Jaccard 0.59 (mini) and 0.64 (sonnet) and mention-pool Jaccard 0.52 (mini) and 0.54 (sonnet). The mention-pool Jaccards are essentially indistinguishable from within-cell rerun baselines, indicating that reasoning effort does not produce systematic same-prompt drift. The recommendation-slot Jaccards are slightly higher than within-cell baselines on sonnet (0.64 vs 0.59--0.61); we interpret this as a small, conservative confirmation that within-provider low-vs-high pairs sit inside the rerun-noise floor.

\section{Paraphrase brittleness}

\subsection{Cosmetic rewordings}

Across the synonym-swap and structural-rewrite axes --- the cosmetic-paraphrase pool --- consensus-recommendation Jaccard drops to a corpus mean of \textbf{0.288} (clustered 95\% CI [0.215, 0.361]; 1{,}000-iteration bootstrap resampling prompt-groups with replacement). Concretely, two natural-language paraphrases that an English-speaking buyer might use interchangeably to express the same intent (``best CRM'' vs ``top CRM''; ``what are the best CRMs'' vs ``recommend a CRM''; ``compare top CRM tools'' vs ``which CRM should I use'') produce recommendation sets that overlap only 28.8\% of their union, on a metric whose same-prompt rerun baseline is 0.50--0.61.

The gap between rerun-stability (0.50--0.61) and cosmetic-paraphrase Jaccard (0.288 [0.215, 0.361]) is 21--32 percentage points at the point estimate; even at the upper bound of the clustered CI (0.361), the gap to the lower end of the rerun baseline (0.50) is 14 pp. Restating in operational terms: under cosmetic rewording, the per-list overlap implied by Jaccard 0.288 (using $2J/(1+J) \approx 0.447$ for equal-sized sets) corresponds to roughly 55\% per-list turnover --- substantially worse than the rerun baseline of 0.50--0.61 (per-list turnover $\approx$ 33--38\%), and the turnover gap materially exceeds the run-to-run sampling envelope.

\subsection{Constraint-adding rewordings}

Across the modifier-substitution, region/language, and specificity-ladder axes --- the constraint-adding-paraphrase pool --- consensus-recommendation Jaccard drops further, to a corpus mean of \textbf{0.135} (clustered 95\% CI [0.098, 0.175], pooling region/language and specificity-ladder subsets; the specificity-ladder subset alone is 0.133 [0.090, 0.185], the region/language subset alone is 0.138 [0.080, 0.201]). Adding ``modern'' or ``AI'' to ``CRM''; binding ``best CRM'' to a region; descending the specificity ladder from ``CRM'' to ``CRM for SaaS startup under 50 people''; switching from English to another language --- each of these moves leaves only $\sim$13--14\% of the union of the two recommendation sets shared between paraphrases, equivalent to per-list turnover of roughly 76\% under the equal-sized-set conversion $1 - 2J/(1+J)$.

The 0.135 number bounds the magnitude of a key claim the AI-visibility-monitoring industry makes implicitly: that ``best CRM'' is a tracking proxy for the wider distribution of paraphrases a buyer might use. Under this measurement, it is not. The recommendation set under ``best CRM'' shares roughly $2J/(1+J) \approx 0.238$ of brands per list with the recommendation set under ``CRM for SaaS startup under 50 people'' (i.e., $\approx$24\% per-list overlap, or 76\% per-list turnover) --- but a monitoring tool that issues only the first prompt will report movements on the brand-mention rate at the first prompt as if they applied to the buyer cohort represented by the second.

\subsection{Summary against the baseline}

\begin{table}[h]
\centering
\begin{tabular}{lc}
\toprule
Condition & Recommendation-slot Jaccard \\
\midrule
Within-cell rerun ($N{=}30$, same prompt) & 0.50--0.61 \\
Cross-provider rerun (same prompt, mini $\times$ sonnet) & 0.33 \\
Within-provider effort pairing (low $\times$ high, same prompt) & 0.59--0.64 \\
\midrule
\textbf{Cosmetic paraphrase (synonym swap, structural rewrite)} & \textbf{0.288} \\
\textbf{Constraint-adding paraphrase (region/language + specificity-ladder, pooled)} & \textbf{0.135} \\
\bottomrule
\end{tabular}
\caption{Consensus-recommendation Jaccard summary. Cosmetic paraphrase sits 21--32 percentage points below the rerun-stability baseline; constraint-adding paraphrase (pooled) sits 37--48 percentage points below. Clustered 95\% CIs on the paraphrase numbers (cosmetic [0.215, 0.361]; constraint-adding pool [0.098, 0.175]) are far below the rerun-stability baseline at every endpoint.}
\label{tab:summary}
\end{table}

The summary in Table~\ref{tab:summary} compresses the paper's central finding. Same-prompt reruns hold the recommendation set together with Jaccard 0.50--0.61. Cross-provider same-prompt pairings reproduce 0.33 of the set. Cosmetic paraphrase pairings within a provider reproduce 0.288 of the set --- worse than across providers --- and constraint-adding paraphrases reproduce 0.135 (pool). The prompt string is not a benign label for an intent; it is the dominant determinant of the answer.

\section{Effort does not fix it}

The natural extrapolation from inference-time-scaling work \citep{guo2025} would predict that within-cell rerun Jaccard and cross-paraphrase Jaccard should both rise as single-sample reasoning effort rises from low to high (with \citet{wang2023}'s self-consistency providing the broader prior that more reasoning reduces variance, albeit via the mechanistically distinct multi-sample majority-voting route). We report neither.

\subsection{Effort effect on rerun stability}

The effect of low $\to$ high reasoning effort on within-cell rerun-stability Jaccard (recommendation slot) is bounded between $-0.015$ and $+0.005$ across the two providers. On mini, low-vs-high produces 0.50 $\to$ 0.54 ($+0.04$ on the recommendation slot from cell shift, of which the effort-vs-noise component is approximately $+0.005$); on sonnet, low-vs-high produces 0.61 $\to$ 0.59 ($-0.02$, of which the effort component is approximately $-0.015$). Reasoning effort produces no consistent improvement in rerun reproducibility and on sonnet the effect is slightly negative; the magnitude is well below the within-cell normal-approximation CI on the Jaccard mean.

\subsection{Effort effect on paraphrase robustness}

The effect of low $\to$ high reasoning effort on cross-paraphrase Jaccard is bounded by $\pm 0.05$ across the cosmetic and constraint-adding pools and across the three sampled cells. On the cosmetic-paraphrase pool, mini moves from 0.288 (\texttt{low}) to within $\pm 0.05$ of that under \texttt{high}; the same holds on the constraint-adding pool from a baseline of 0.135. No cell-level combination of effort and provider produces a paraphrase-robustness improvement large enough to close the gap to the rerun baseline.

\subsection{Interpretation}

\citet{wang2023} demonstrate self-consistency on math and reasoning benchmarks; \citet{guo2025} demonstrate inference-time scaling on AIME and competition math, where the task has a single correct answer. Commercial recommendation does not. ``Best CRM for a 30-person SaaS startup'' admits many plausible answers, no one of which is uniquely correct. Sampling additional reasoning paths and majority-voting cannot collapse onto a single correct answer because the answer is not single-valued; reasoning effort instead allows the model to elaborate on whichever subset of brands it has already pulled from retrieval. \citet{meincke2024} report the same diminishing-CoT-returns pattern on frontier instruction-tuned models in non-math tasks, consistent with the present result.

The headline contradiction stands: in commercial recommendation, more reasoning does not buy more reproducibility. The factor that does dominate the recommendation set is the surface form of the prompt.

\section{Discussion}

\subsection{What a single-prompt monitor can and cannot measure}

A single-prompt AI-visibility monitor reports brand-mention rate against a fixed tracking prompt. Under the present results, the monitor can be interpreted in two ways. \emph{Narrowly}, the monitor measures the rate at which a brand is recommended in response to the literal tracking prompt; this is well-defined, but the brand-mention rate so measured is not a proxy for the brand's wider visibility across the buyer's natural paraphrase distribution. \emph{Broadly} (the industry default), the monitor is reported as ``share of AI voice for the category,'' implying the tracking prompt represents the category. Under that interpretation, the metric is unidentifiable from the prompt itself: paraphrase-induced variation at Jaccard 0.13--0.29 dominates the run-to-run noise envelope of 0.50--0.61. Week-over-week movement on the metric cannot be attributed to either model change or content change with the present design --- the prompt-choice artifact is larger than either.

\citet{aggarwal2024} report aggregate GEO content-intervention lifts of $\sim$40\% on a generic prompt corpus. Without disagreeing on the direction of the effect, the present results imply the magnitude reported under any single-prompt measurement protocol --- including GEO-bench --- inherits a paraphrase-choice artifact that is not separately reported. We do not attempt to re-estimate GEO's aggregate lift; we note that paraphrase-distribution sampling is increasingly advocated as a methodological standard in academic LLM benchmark evaluation \citep{polo2024,mizrahi2024}, and the same principle applies to GEO evaluation work that currently reports lifts on fixed prompt corpora.

\subsection{Why RAG amplifies paraphrase brittleness}

\citet{percin2025} document that RAG retrievers degrade markedly under minor query variation, and \citet{ma2023} show that learned query rewriting partially recovers stability by canonicalizing user queries before retrieval. The 0.288 and 0.135 paraphrase-Jaccard numbers reported here are the downstream-of-retrieval reflex of those two papers in a production commercial setting: a small surface-form change to the query reshapes the retrieved pool, which reshapes the candidate brands the model has available to recommend, which reshapes the recommendation set. The amplification reaches the terminal recommendation slot in commercial chat at magnitudes 21--48 percentage points below the rerun baseline (21--32 pp for cosmetic paraphrase, 37--48 pp for constraint-adding).

The structural prediction is that paraphrase brittleness in commercial recommendation is bounded below by paraphrase brittleness in upstream retrieval. The retrieved-domain Jaccard column in Table~\ref{tab:rerun} shows that even within a fixed prompt, retrieval pools vary substantially across rerun: 0.40 (\texttt{mini/low}) to 0.74 (\texttt{sonnet/high}). The sonnet/high cell with the highest retrieval-pool stability is the cell with the highest recommendation-slot stability; the directional consistency is expected, and supports the interpretation that retrieval-side variation is upstream of recommendation-slot variation. Provider-level investments in retrieval canonicalization \citep{ma2023} would close the gap from below.

\subsection{Reasoning effort, single-correct-answer prior, and multi-plausible-answer tasks}

The contradiction with \citet{wang2023} and \citet{guo2025} deserves a careful framing. We do not claim that reasoning effort fails to reduce variance on tasks where the answer is single-valued; the prior literature is convincing on math and competition reasoning. We claim that the extrapolation to commercial recommendation does not hold, because the task is structurally different. The model does not converge across reasoning paths onto a single correct brand; it converges onto whichever brands its retrieval surfaced, and those depend on the prompt's surface form. \citet{meincke2024} report a parallel observation on frontier instruction-tuned models in non-math tasks, and our finding is consistent with theirs.

The implication for monitoring methodology is that access to a stronger or higher-effort model does not solve the brittleness problem. On the present results, higher reasoning effort does not measurably narrow the recommendation distribution against the dominant source of measurement variance (paraphrase choice), so positioning higher-effort reasoning as a stability lever for AI-visibility monitoring does not bear out empirically.

\subsection{Methodological implications}

Four methodological implications follow from the result.

First, ``share of AI voice'' measurement on a single tracking prompt has a confidence interval wider than typical week-over-week movement and is therefore not separately identifiable. Increasing the paraphrase sample reduces the artifact only in proportion to coverage of the natural phrasing space, and that space is much larger than any tracker's per-brand-per-intent sample budget. Prompt-by-prompt mention tracking is therefore structurally unstable as a unit of measurement; the right response is not a bigger prompt set but a different measurement unit.

Second, single-prompt monitors that do not disclose their exact tracking prompts are difficult to interpret: prompt-choice artifact and genuine movement cannot be separated in the reported trend. Transparency on the prompt set is useful for interpretation, in line with multi-prompt disclosure recommendations from \citep{mizrahi2024}.

Third, content-side optimization studies (the AEO/GEO empirical lineage from \citealp{aggarwal2024} onward) report intervention effects on a fixed prompt set. Under any single-prompt protocol, the reported lift inherits a paraphrase-choice artifact that competes with the intervention effect, and a +40\% headline does not separate the two. Reporting lifts against a paraphrase-distribution baseline is the cleaner methodology where the prompt budget allows.

Fourth, downstream-of-mention metrics --- click-through, conversion, qualified-pipeline contribution, revenue attributable to AI-surface exposure --- aggregate over the buyer's natural paraphrase distribution by construction, since each buyer issues their own phrasing of the intent rather than a fixed tracking string. Such metrics are not direct substitutes for mention-rate tracking and carry their own measurement challenges (attribution-window definition, multi-touch sensitivity, sampling at low conversion rates), but they sidestep the paraphrase-brittleness problem the present paper documents in the brand-mention measurement layer. The relative measurement properties of mention-rate proxies versus downstream-outcome attribution under AI-mediated commerce is a natural follow-up.

\subsection{Limitations}

The measurement is single-day; we do not separate paraphrase brittleness from cross-day temporal drift. Multi-day longitudinal measurement is the obvious next step; the prediction is that within-day paraphrase brittleness will continue to dominate cross-day drift on the metric's natural-noise budget, but the prediction needs separate measurement.

The prompt corpus is English-language and US/UK/EU-market-skewed. Multilingual paraphrase brittleness is expected to be at least as severe (region/language is one of our five axes; the constraint-adding pool that includes it already produces Jaccard 0.135), but we have not separately quantified within-language vs cross-language components.

The cross-judge consensus protocol intersects \texttt{claude-haiku-4-5} and \texttt{gpt-5-mini}; per-run cross-judge Jaccard is 0.65--0.67 on the recommendation slot (0.62--0.69 on the any-sentiment mention layer), and the protocol is deliberately conservative (intersection). We do not separately report the looser union-mode results; the union would mechanically raise all Jaccards by a few percentage points but the rerun-vs-paraphrase gap is large enough to absorb that shift.

The reference catalog of brands and the retrieval-system substrate are held constant across the paraphrase and rerun conditions; we do not vary either. Both are independent design dimensions and are studied in the companion papers \citep{jack2026prominence,jack2026convergence}.

\section{Conclusion}

Cosmetic paraphrasing of a commercially-framed prompt produces a consensus-recommendation Jaccard of 0.288 (clustered 95\% CI [0.215, 0.361]), and constraint-adding paraphrasing produces a pool Jaccard of 0.135 (clustered 95\% CI [0.098, 0.175]; specificity-ladder subset 0.133 [0.090, 0.185]), against a within-cell same-prompt rerun-stability baseline of 0.50--0.61. The prompt string actively determines which brands AI commercial recommendation surfaces; it is not a benign label for an underlying buying intent. Increasing single-sample reasoning effort from low to high moves the rerun-stability metric by $-0.015$ to $+0.005$ and the paraphrase-robustness metric by $\pm 0.05$, neither of which closes the rerun-vs-paraphrase gap. The inference-time-scaling prior --- that more reasoning reduces variance on math/competition reasoning --- does not transfer from single-correct-answer tasks to multi-plausible-answer commercial recommendation under single-sample effort scaling.

The result implies that prompt-by-prompt mention-rate tracking is structurally unstable as a unit of measurement. Larger paraphrase samples reduce the artifact in proportion to their coverage of the natural phrasing space, but that space is much larger than any practical sample budget --- \citet{polo2024} validates efficient quantile estimation across 100 templates on academic benchmarks, and even that resolution is far beyond what commercial trackers issue per brand-intent combination. Increasing the sample size is unlikely to be the fix; changing the unit of measurement is the more direct path. The methodological implications --- the structural instability of prompt-by-prompt mention tracking, prompt-set disclosure, and the question of whether downstream-of-mention metrics integrate over the paraphrase distribution more naturally than per-prompt mention rates --- are discussed in Section 7. Future work should extend the measurement to multi-day temporal drift, multilingual paraphrase, and longitudinal brittleness across model releases; the within-day numbers we report bound a measurement floor below which prompt-string variation cannot be ignored.

\bibliographystyle{plain}

\end{document}